\documentclass[12pt, a4paper]{article}

\def\bea{\begin{eqnarray*}}
\def\eea{\end{eqnarray*}}
\def\bean{\begin{eqnarray}}
\def\eean{\end{eqnarray}}
\def\bi{\begin{itemize}}
\def\ei{\end{itemize}}

\let\proglang=\textsf

\def\cmin{{C_\text{min}}}

\usepackage{amsmath}
\usepackage{amssymb}
\usepackage{amsfonts}
\usepackage{amsthm}
\usepackage{latexsym}
\usepackage{color}
\usepackage{epsfig}
\usepackage{hyperref}
\usepackage{longtable}
\usepackage{multirow}
\usepackage{lscape}
\usepackage[normalem]{ulem}   % for use of \sout
\usepackage{ulem}

\newcommand{\RR}{\text{RR}}

\begin{document}

\title{Principal Stratum Strategy: Potential Role in Drug Development}

\author{
    Bj\"orn Bornkamp\thanks{Clinical Development and Analytics, Novartis, Basel, Switzerland} \\
    Kaspar Rufibach\thanks{Methods, Collaboration, and Outreach Group (MCO), Department of Biostatistics, Hoffmann-La Roche Ltd, Basel, Switzerland} \\
    Jianchang Lin\thanks{Statistical \& Quantitative Sciences (SQS), Takeda Pharmaceuticalsw, Cambridge, MA, USA} \\
    Yi Liu\thanks{Nektar Therapeutics, San Francisco, CA, USA} \\
    Devan V. Mehrotra\thanks{Merck \& Co., Inc., North Wales, PA, USA} \\
    Satrajit Roychoudhury\thanks{Pfizer Inc., New York, NY, USA} \\
    Heinz Schmidli\footnotemark[1]\\
    Yue Shentu\thanks{Merck \& Co., Inc., Rahway, NJ, USA}\\
    Marcel Wolbers\footnotemark[2] \\
}

\date{\today}

\maketitle

\begin{abstract}
  %% Needed to shorten abstract (Pharm. Stat only allows 250 words)
  A randomized trial allows estimation of the causal effect of an
  intervention compared to a control in the overall population and in
  subpopulations defined by baseline characteristics. Often, however,
  clinical questions also arise regarding the treatment effect in
  subpopulations of patients, which would experience clinical or
  disease related events post-randomization. Events that occur after
  treatment initiation and potentially affect the interpretation or
  the existence of the measurements are called {\it intercurrent
    events} in the ICH E9(R1) guideline. If the intercurrent event is
  a consequence of treatment, randomization alone is no longer
  sufficient to meaningfully estimate the treatment effect. Analyses
  comparing the subgroups of patients without the intercurrent events
  for intervention and control will not estimate a causal effect.
  This is well known, but post-hoc analyses of this kind are commonly
  performed in drug development. An alternative approach is the
  principal stratum strategy, which classifies subjects according to
  their potential occurrence of an intercurrent event on both study
  arms. We illustrate with examples that questions formulated through
  principal strata occur naturally in drug development and argue that
  approaching these questions with the ICH E9(R1) estimand framework
  has the potential to lead to more transparent assumptions as well as
  more adequate analyses and conclusions. In addition, we provide an
  overview of assumptions required for estimation of effects in
  principal strata. Most of these assumptions are unverifiable and
  should hence be based on solid scientific understanding. Sensitivity
  analyses are needed to assess robustness of conclusions.

\end{abstract}

%% Shortened keywords to 5 (as required by Pharm Stat)

% % -----------------------------------------------------
% \section*{Stuff \& to do}
% % -----------------------------------------------------
%
% \bi
% \item Targeted journal: Pharmaceutical Statistics?
% \item Github repository: \url{https://github.com/rufi77/princ_strat.git}, username: rufi77, login: miliflue18
% \item Heterogeneous use of posttreatment, post-baseline, post-randomization. Align?
% \ei

% -----------------------------------------------------
\section{Introduction}
\label{intro}
% -----------------------------------------------------

One main concept of the E9(R1) guideline \cite{ich_19} by the
International Council of Harmonization (ICH) is the notion of
intercurrent events, defined as ``... Events occurring after treatment
initiation that affect either the interpretation or the existence of
the measurements associated with the clinical question of
interest. ...''.  The ICH E9(R1) guideline outlines five strategies to
acknowledge intercurrent events as part of the treatment
effect/estimand of interest. The treatment policy strategy effectively
makes the intercurrent event part of the treatment investigated. The
composite and while-on-treatment strategies modify the
variable/endpoint of interest to reflect the intercurrent event. The
hypothetical strategy envisages a hypothetical scenario in which the
intercurrent event does not occur.  Finally, the principal stratum
strategy, based on ideas introduced by \cite{frangakis_02}, defines a
subpopulation of interest according to the potential occurrence of an
intercurrent event on one or all treatments.

As part of a principal stratum strategy, the subpopulation of interest
could for example be subjects who \textit{would tolerate treatment if
  assigned to the test treatment}. In this case subpopulation
membership on the test arm would be known. On the control arm however
subpopulation membership is not observed and hence not known with
certainty. Alternatively, the subpopulation of interest could be the
patients who would tolerate both test and control treatment.

The principal stratum strategy has not been commonly used in clinical
trials so far and is not uncontested, see \cite{hernan_18,
  scharfstein_19} or the discussion initiated earlier by
\cite{pearl_11}.  First, it relates to a subpopulation of the overall
trial population that is not identifiable with certainty
(\textit{i.e.} for some, or all, patients principal stratum membership
is not observed and constitutes missing data). This may be perceived
to render the obtained treatment effect estimate of limited interest
from a direct practical perspective. Second, a principal stratum
estimand relates to a question where one cannot rely on randomization
anymore to ensure comparable baseline populations across treatment
groups in the subpopulation of interest.  Strong assumptions are
typically needed to estimate this estimand. In the ICH E9(R1)
guideline it is specifically mentioned that a run-in period may be an
effective design feature to robustly identify a target population
defined by a specific clinical event (and thus estimate a principal
stratum effect). The use of these designs might however be limited to
special situations.
% \sout{Further complications arise in particular for
% time-to-event type endpoints: occurence of the primary event might
% make observation of the intercurrent event status impossible.  Naive
% analyses conditioning on observed intercurrent event occurence would
% then not only compare non-randomized populations but may also suffer
% from immortal bias, as some of the patients with early events might
% have had the intercurrent event later, i.e. those for which we observe
% the intercurrent event are ``immortal'' until that
% timepoint.}\bb{Propose to delete this here as this is mentioned in
% Section 2 and then in 4.7 in more detail.}

For these reasons, one might be tempted to generally challenge the
relevance of the principal stratum strategy in drug development.

In this paper we would like to illustrate with examples that many
relevant scientific questions in drug development can be
addressed with the principal stratum strategy. Often, these questions
do not correspond to the primary endpoint in the specific trial, but
they increase the scientific understanding of the treatment effect in
relevant subpopulations, and may impact approval decisions and
labeling.

The outline of this paper is as follows.  In Section~\ref{sec:pot_out}
we will provide a review of potential outcomes and principal stratum
estimands.  In Section~\ref{examples} we review examples from drug
development practice, where the question of interest can be framed to
be of principal stratum type. Section~\ref{sec:analysis} then reviews
analysis methods and assumptions citing existing literature and
outlines a \proglang{R} \cite{r} implementation. The paper ends with a
discussion in Section~\ref{discussion}.

% -----------------------------------------------------
\section{Introduction to Potential Outcomes and Principal Stratum Estimands}
\label{sec:pot_out}
% -----------------------------------------------------

The term {\it principal stratum} was first introduced by
\cite{frangakis_02} (see also \cite{mealli_12} for a rather recent
review on principal stratification) and originates from the causal
inference literature under the potential outcome approach (see
\cite{hernan_20, imbens_15} for introductions). In this section we
will introduce potential outcomes, a central idea of causal inference,
which are important to formulate principal stratum estimands. Note
that the other estimand strategies in the ICH E9(R1) guideline can
also be formulated using potential outcomes \cite{lipkovich_20}. We
illustrate potential outcomes with an example: Let $Z$ be the binary
indicator for treatment ($Z=1$ corresponding to the test treatment and
$Z=0$ corresponding to control) and $Y$ be the outcome of
interest. Assume a treating physician is deciding on the treatment to
prescribe. Ideally she would make that decision based on knowledge on
what the outcome for the patient would be if given the control
treatment, $Y(Z=0)$, abbreviated as $Y(0)$, and what the outcome would
be under test treatment, $Y(Z=1)=Y(1)$. In reality of course, neither
$Y(0)$ and $Y(1)$ is known when assigning a treatment, and even after
observation, for a given patient, only one of the potential outcomes
$Y(0)$ or $Y(1)$ can be observed. So, even after observation of $Y$
one cannot be sure if the correct decision was made for this
particular patient: Individual causal effects, \textit{i.e.}
$Y(1)-Y(0)$, are not observed. On a population level, however, such
``causal'' statements can be made. One then targets the average causal
effect $E(Y(1)-Y(0))$, where the expectation is taken with respect to
the population of interest.

% \sout{Statistical estimation of $E(Y(1)-Y(0))$ in a randomized trial
%   can be performed based on the fact that the population on the
%   intervention arm is exchangeable with the population on the control
%   arm (and thus also with the overall population). So we have that
%   $Y(1)$ and $Y(0)$ are independent of $Z$ implying that}

Statistical estimation of $E(Y(1)-Y(0))$ in a randomized trial can be
performed based on the fact that treatment assignment is independent
of any patient characteristic, so that $Y(1)$ and $Y(0)$ are
independent of $Z$ implying that
\bea
  E(Y(1)-Y(0)) &=& E(Y(1))-E(Y(0)) \\
  &=& E(Y(1)|Z=1) - E(Y(0)|Z=0) \\
  &=& E(Y|Z=1) - E(Y|Z=0).
\eea
  This means we can estimate the average causal effect by the
  difference in averages on the two arms, as the population of
  patients is comparable across the two treatment arms.

%   \sout{Two groups are {\it exchangeable} with respect to an outcome measure if their outcomes would be the same whenever they were subjected to the same treatment. Exchangeability is e.g. fulfilled in a randomized controlled trial (RCT).}
% \bb{Something wrong in the sentence above?}\kr{Better now?}\bb{I am
%   not sure this is the definition of exchangeability. I propose to
%   remove the text here and re-wrote the part above without the term
%   exchangeability. Ok like this?}

In an observational study the treatment decision between $Z=0$ and $Z=1$
might depend on further measured or unmeasured patient characteristics
$X$, so that the patients who receive $Z=1$ (for whom we observe $Y(1)$)
might be systematically different from those patients who receive
$Z=0$ (for whom we observe $Y(0)$), so that $Y(1)$ and $Y(0)$ are not
independent of $Z$. In this case
$E(Y|Z=1) - E(Y|Z=0) \neq E(Y(1)-Y(0))$, because
$E(Y(1)) \neq E(Y(1)|Z=1)$ and $E(Y(0)) \neq E(Y(0)|Z=0)$: The
patients receiving $Z=0$ are not representative of the overall
population and similarly those receiving $Z=1$ are not representative
of the overall population. The value of potential outcomes from a
notational perspective is that they allow to decouple the outcome
$Y$ from the actual treatment $Z$ received.

\begin{figure}[b!]
  \begin{center}
    \includegraphics[width=0.95\textwidth]{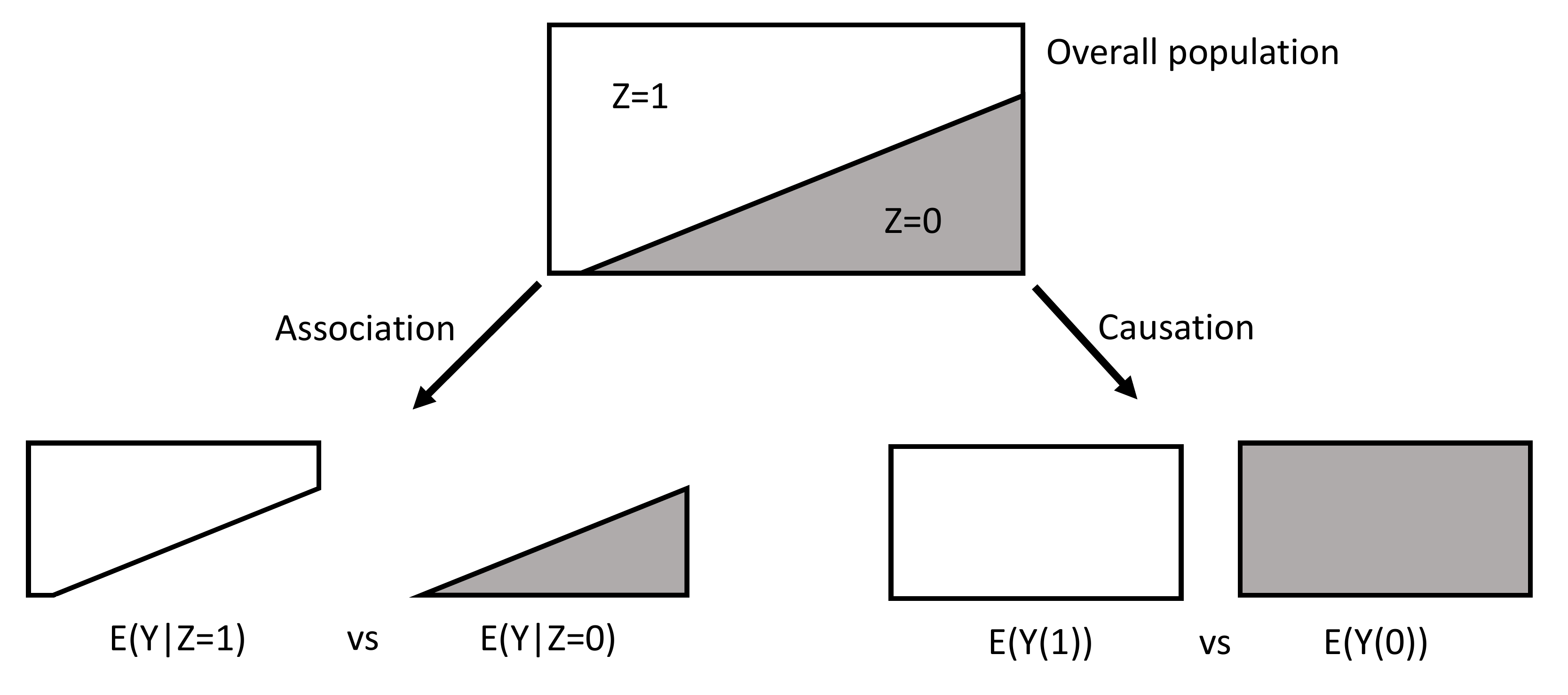}
  \end{center}
  \caption{Difference between association and causation (adapted based
    on \cite[Figure 1.1]{hernan_20})}
  \label{fig:assoc_vs_causal}
\end{figure}

Denoting by $Y(1)_i$ the potential outcome for a patient $i$ and by
$\mathcal{S}$ a population of patients, causal treatment effects are
defined as a comparison of potential outcomes
$\{Y(1)_i, i\in \mathcal{S}\}$ versus $\{Y(0)_i, i\in \mathcal{S}\}$
on a common set of units $\mathcal{S}$. A comparison of
$\{Y(1)_i, i\in \mathcal{S}_1\}$ versus
$\{Y(0)_i, i\in \mathcal{S}_2\}$ with
$\mathcal{S}_1 \neq \mathcal{S}_2$ is not a causal effect
\cite{rubin_05}. A causal effect can thus be conceptualized as a
comparison of outcomes ``had everyone received treatment'' versus
outcomes ``had everyone received control'', see also Figure
\ref{fig:assoc_vs_causal}. This focus on causal effects is also
present in the ICH E9(R1) guideline \cite[Section A.3]{ich_19}, where
estimands are introduced as ``... Central questions for drug
development and licensing are to establish the existence, and to
estimate the magnitude, of treatment effects: how the outcome of
treatment compares to what would have happened to the same subjects
under alternative treatment (i.e. had they not received the treatment,
or had they received a different treatment). ...''.

To illustrate a main motivation of the principal stratum strategy
we will consider a simple, generic example. Assume a randomized
two-arm trial is planned, with an outcome $Y$ assessed at week 12. Now
assume that one is interested in the treatment effect in those
patients that experience a specific post-randomization event of
interest. Denote by $S=1$ occurrence and by $S=0$ absence of the
post-randomization event.  A naive analysis that might be employed in
such situations is to subset the overall trial data to patients with
$S=1$ on both test and control arm and then perform the analysis of
interest.  The variable $S$ is a post-randomization variable and an
outcome influenced by treatment, \textit{i.e.}  $S$ depends on
$Z$. This means for patients on the intervention arm we observe the
potential outcome $S(Z=1)$ and on control we observe the potential
outcome $S(Z=0)$.  From this perspective the population of patients
with $S(1)=1$ and $S(0)=1$ might be quite different. The naive
analysis mentioned above is hence ``breaking the randomization'', as
the patient populations on the compared arms can be different, and one
is not comparing ``like with like'', and thus not estimating a causal
effect. If we would numerically observe a treatment effect in such an
analysis, we would not be sure whether the difference in outcome is
due to the difference in treatment, or due to the difference in the
compared populations.

The idea of principal stratum estimands is to stratify patients based
on their potential outcomes $S(0),S(1)$ for {\it all} treatments. In
the case of a binary post-randomization event $S$ and two treatments
one can hence define four strata based on both potential outcomes, see
Table~\ref{tab:principal_stratum}. Every patient falls into one
particular of the four strata. Causal interpretations are made possible by the fact
  that membership to a principal stratum is not affected by treatment
  assignment.

\begin{table}[t]
  \centering
  \begin{tabular}{c|c|c}
               & $S(0) = 1$                   & $S(0) = 0$                   \\ \hline
    $S(1) = 1$ & $\{S(1) = 1\} \cap \{S(0) = 1\}$ & $\{S(1) = 1\} \cap \{S(0) = 0\}$ \\ \hline
    $S(1) = 0$ & $\{S(1) = 0\} \cap \{S(0) = 1\}$ & $\{S(1) = 0\} \cap \{S(0) = 0\}$ \\
  \end{tabular}
  \caption{Principal strata defined by the potential outcomes $S(0)$ and $S(1)$.}
  \label{tab:principal_stratum}
\end{table}

In the described setting, patients that would experience the event
under either treatment would have $S(1)=1$ and $S(0)=1$ (\textit{i.e.}
the top-left cell in Table \ref{tab:principal_stratum}).
Alternatively of interest could be those patients that experience the
event under treatment $S(1)=1$, which is the union of the two cells
$\{S(1)=1\} \cap \{S(0)=1\}$ and $\{S(1)=1\} \cap \{S(0)=0\}$ in Table
\ref{tab:principal_stratum}.

Contrary to the naive analysis this stratification leads to a causal
effect: We now stratify the population according to the same rule on
treatment and control arm. 
% \sout{Although within the potential outcomes framework it is defined
%   at baseline to which stratum a patient belongs,}\bb{Propose to
%   remove this} 
  The actual {\it identification} of the subpopulation
corresponding to the stratum/strata of interest is generally not
possible, not even after observing the outcome $Y$ and the
post-randomization event $S$ in a given trial. For patients on the
intervention arm we observe $S(1)$, but not $S(0)$, and vice versa for
patients on the control arm.

Based on our experience (see also the examples in Section
\ref{examples}) one is often interested in the group of patients that
experience the post-randomization event under one treatment (union of
two principal strata, or a row in Table~\ref{tab:principal_stratum}),
for example the stratum with $S(1)=1$. Then in one arm patients in the
stratum can be identified, but not on the other arm.  Generally,
assumptions are required for estimation of the treatment effect in the
stratum/strata of interest. One could argue that the naive analysis
(that subsets based on the observed $S$ on both treatment arms) is
estimating the treatment effect in the principal stratum
$\{S(1)=1\} \cap \{S(0)=1\}$ under the assumption that $S(1)=S(0)$,
\textit{i.e.}  occurrence of the post-randomization event is not
treatment related. Viewing this naive analysis within potential
outcome notation reveals the implicit assumption underlying the
analysis, which might often be quite strong and rarely justified.

While often one is primarily
  interested only in a subset of the overall trial population (one
  principal stratum or a union of strata), it is good practice to
  evaluate and report results also for the complementary group(s) or
  all strata (or union of strata) if the model allows to extract that
  information.

  A complication arises when the outcome $Y$ is not a measurement
  assessed at a specific timepoint, but of time-to-event type, such as
  death for overall survival (OS). In this situation the event itself
  can constitute a competing risk for occurrence of the intercurrent
  event (\textit{i.e.}  after observing the main event of interest,
  patients would no longer be at risk for experiencing the
  intercurrent event). In these situations particular care is needed
  to define the principal stratum of interest as well as the analysis
  strategy. Naive analyses conditioning on observed intercurrent event
  occurrence in this situation would then not only compare
  non-randomized populations but may also suffer from immortal bias,
  as patients for which we observe the intercurrent event are
  ``immortal'' until that timepoint.

  In the causal inference literature the principal stratum approach
  has been controversially discussed in particular with respect to its
  relationship to mediation analysis, see for example
  \cite{pearl_11}. In the latter, one tries to disentangle the overall
  effect via direct and indirect effects (mediated via the
  intercurrent event or not).  In the language of the ICH E9(R1),
  mediation analysis can be interpreted as targeting a hypothetical
  estimand. See also the comparisons between principal stratum and
  mediation approaches in \cite{vanderweele_08} on a conceptual level,
  and \cite{baccini_17} on a practical example. In this paper, we will
  focus on principal stratum estimands, not the least because this
  concept has been proposed as one of five strategies to address
  intercurrent events in the ICH E9(R1) guideline.

\section{Examples}
\label{examples}
% -----------------------------------------------------

% \bi
% \item BAF (Björn) Explain situation, principal strata, estimand and analysis (maybe add code?)
% \item early biomarker (Björn) Explain situation, principal strata, estimand and analysis (maybe add code?)
% \item 2 Examples around tumor shrinkage (Kaspar)
% \item Low vs high exposure (Kaspar)
% \item ADA Example (Kaspar)
% \item Switching to higher dose (Kelly)
% \item CAR-T bridging chemo (Björn)
% \ei
%
% \kr{I have extended example sections below, especially Section~\ref{ex_ada}. I somehow told the early biomarker story within that section, it appears...}

In this section we discuss scientific questions of interest in drug
development that are formulated as principal stratum estimands. For a
discussion how to position these questions in the broader drug
development landscape we refer to Section~\ref{discussion}. In this
section we will not discuss analysis strategies or assumptions that
would allow estimation. We return to this aspect in
Section~\ref{sec:analysis}. Table~\ref{tab:examples} gives an overview
of the examples.

\newpage

\begin{landscape}
\renewcommand{\baselinestretch}{1}
\begin{table}[h]
\begin{center}
\begin{tabular}{p{4cm}|p{6.5cm}|p{3.5cm}|p{4cm}|p{4cm}}\hline
Example            & Scientific question                              & Primary endpoint & Intercurrent event & Stratum of interest          \\ \hline
Multiple Sclerosis & Treatment effect on confirmed disability progression in the subpopulation of relapse-free patients & Time to confirmed disability progression & Post-randomization relapse  & Patients who would be relapse-free under both treatments\\ \hline 
Treatment effect in early responders & Predict treatment effect on long-term primary endpoint based on early biomarker-type readout & Time-to-event & Biomarker value above or below a pre-specified threshold & Patients who would respond early under treatment vs. those that would not  \\ \hline
Antidrug antibodies (ADA) for targeted oncology drugs & Do patients that develop ADAs on either arm still benefit from the drug? & Time-to-event & Development of antidrug antibodies because of receiving treatment & Patients who would be ADA+ under treatment \\ \hline
Impact of exposure on OS & Do patients with insufficient exposure have lower treatment effect? & Time-to-event & Exposure below a pre-specified threshold & Patients with low vs. non-low exposure under treatment\\ \hline
Prostate cancer prevention & Assess effect of treatment to prevent prostate cancer on severity of prostate cancer among those men who would be diagnosed
with prostate cancer regardless of their treatment assignment & Time-to-event & Getting prostate cancer & Patients who get prostate cancer irrespective of treatment \\ \hline
\end{tabular}
\caption{Examples of principal stratum estimands discussed in this section.}
\label{tab:examples}
\end{center}
\end{table}
\renewcommand{\baselinestretch}{1.0}
\end{landscape}

\newpage

% -----------------------------------------------------
\subsection{Multiple sclerosis}
\label{ms_example}
% -----------------------------------------------------

Multiple sclerosis (MS) is an auto-immune disease of the central
nervous system characterized by relapses, with varying symptoms, for
example visual deficits, cognitive and motor impairment. Multiple
sclerosis typically starts out with a phase where patients have
relapses, but fully recover after the relapses (relapsing remitting
form of MS, RRMS). Then the disease transitions to a phase where
patients have a continuous disease progression where relapses are less
common, and patients often do not fully recover from these, leading to
increased disability (secondary progressive MS, SPMS). The typical
primary endpoint in RCTs for SPMS is time to confirmed disability
progression. In Magnusson et al. \cite{magnusson_19} EXPAND (NCT01665144,
\cite{kappos_18}), a large placebo-controlled trial of siponimod in
patients with SPMS, is discussed. The primary objective of the trial
was to show efficacy of siponimod versus placebo in terms of time to confirmed
disability progression. The endpoint was achieved, but the question
was raised whether a treatment effect would also be present in patients that would not experience relapses. As siponimod is known to prevent relapses this is a
non-trivial question to answer.

In this setting the intercurrent event $S$ is post-randomization relapse,
and \cite{magnusson_19} considered estimation of the treatment effect
in patients that would not relapse under both siponimod and placebo,
\textit{i.e.} in the stratum $\{S(0) = 0\} \cap \{S(1) = 0\}$.  The probability
of disability progression was assessed at a specific timepoint, so
that the outcome $Y$ is binary. The estimand of interest here was
taken as the risk ratio
\bea
  \RR &:=& \frac{E(Y(1) | S(1) = 0, S(0) = 0)}{E(Y(0) | S(1) = 0, S(0) = 0)}. \label{eq:rr}
\eea

% \bb{Heinz mentioned that for PPMS no relapses are expected, should we
%   discuss anything about that here, or not?}
% \kr{Fine for me not to extend the discussion here, as I consider the point that we want to make made?}
% \kr{Should we say why for the PS analysis, the looked into a dichotomized endpoint? Was that for simplicity?}

% Note that the causal risk ratio in the population is not the average
% of the individual causal effects on the ratio scale. More explicitly,
% it is a measure of causal effect in the population but is not the
% average of any individual causal effects \cite{hernan_20}.

% -----------------------------------------------------
\subsection{Treatment effect in early responders}
\label{early_biomarker}
% -----------------------------------------------------

Biomarkers or early readouts can be useful to investigate whether an
investigational medicine works as intended on a biological level. In
some situations, it is realistic to assume that patients, whose
post-randomization short term biomarker levels indicate that they do not
sufficiently respond to the drug, are also unlikely to respond on
clinically relevant long term outcomes, such as time-to-event.

One recent example is in the cardiovascular area.  Inflammation has
been identified as playing a key role in atherosclerosis and
cardiovascular disease. The CANTOS outcomes trial in prevention of
cardiovascular events (NCT01327846, \cite{ridker_17}) investigated
treatment with canakinumab, an anti-inflammatory agent, against
placebo, both on top of standard of care. The primary outcome was the
time to major adverse cardiovascular event (MACE) and significant. In
this specific case the biomarker of interest is a downstream
inflammatory marker, high sensitivity c-reactive protein (hs-CRP),
where lower values indicate less inflammation. Interest here was in
determination of the treatment effect for patients that, three months
after start of treatment with canakinumab, were able to lower hs-CRP
below a specific target level. As the mechanism of action of
canakinumab is lowering inflammation, one would suspect that patients
who do not achieve the biomarker threshold also have a lower benefit
in terms of the time-to-event outcome. Vice versa, patients that
achieve the threshold have a larger treatment effect.

Another example is in oncology, where tumor size shrinkage
is a measurement that can be assessed early. Again, for patients with a
lack of tumor shrinkage it is less likely that those benefit from the
treatment on longer term survival outcomes. In pharmacometrics
so-called {\it tumor growth inhibition} (TGI) metrics have gained
popularity. Such models are drug-independent and attempt to link tumor
response and baseline prognostic factors to a time-to-event endpoint
such as OS. In these models, tumor response is
quantified by extracting a summary statistics from a longitudinal
model of tumor size. %\cite{han_16} provides an example. 
%Based on data of two large Phase 3 RCTs, a nonlinear mixed effects model was fitted to tumor size over time and TGI metrics were obtained for each
%patient, e.g. ratio of tumor size at week 8 to baseline. 
The goal of these TGI analyses is to ``predict'' OS survival functions
and induced effects of treatment based on such summary statistics (see
for example \cite{han_16}). Again, one could consider the treatment
effect (in terms of OS) in patients that achieve a specific favorable
tumor metric shortly after treatment start.

Determining the potential long-term treatment effect for a patient
based on a short-term read-out, such as e.g. hs-CRP or TGI, can be
useful information: Depending on the therapeutic setting and drug
mechanism it might support the decision on treatment modifications
after treatment start.

Let $S$ denote the event of achieving an early readout value (\textit{i.e.}
hs-CRP or TGI) either (1) lower than a target level or (2) achieving a certain percent
decrease with respect to the patient's baseline value, at a short time
$\tilde{t}$ after start of treatment.

Interest focuses on comparing $Y(1)$ and $Y(0)$ in the stratum of
patients with $S(1)=1$, and contrasting this for example to the
results for patients with $S(1)=0$. Depending on the questions one
could also be interested in the subpopulation of patients with
$S(0)=1$ and contrast results to those with $S(0)=0$. Effect measures of interest can  be based on the survival functions
\bea
U_1(t) \ := \ P(Y(1) > t|S(1) = 1)
&\text{ and }& U_0(t) \ := \ P(Y(0) > t|S(1) = 1),
\eea
e.g. event probabilities at a time $t^* > \tilde{t}$:
\bea
\delta(t^*)&=& U_1(t^*) - U_0(t^*)
\label{eq:surv_diff}
\eea or a time-averaged version
$\int_0^{t^*}\delta(t)dt=E[\min(Y(1), t^*)-\min(Y(0),t^*)]$, the
difference in restricted mean survival times \cite{royston_13}.

An important point to consider in these situations (as discussed in
\cite{anderson_08}) is that response on the early read-out might
simply act as a marker for prognostically favorable patients and thus
not modify the treatment difference versus the control treatment
itself. For example comparing $Y(1)$ for patients with $S(1)=1$ versus
those with $S(1)=0$, which does not allow for a statement on the
treatment effect (which is a contrast involving $Y(1)$ and $Y(0)$).

Another challenge is that, depending on the time point $\tilde{t}$ of
the measurement of the post-randomization marker, some events related
to $Y$ might already have happened. 
% This might mean that patients
% switch to a new treatment and $S$ might not have been measured for
% some patients, leading to a selection problem \sout{as discussed earlier}\kr{Do we really discuss this earlier?}
% (\cite{rosenbaum_84} and immortal bias, see also the discussion in
% Section~\ref{ex_ada}). If $Y$ measures a terminal event like death it
% is more meaningful to consider the population of interest to be the
% population of patients with $S(1)=1, Y(0)>\tilde{t}$ and
% $Y(1)>\tilde{t}$.  While in situations where measurement of $S$ is
% still meaningful even after the event $Y$ has happened one could still
% consider to focus on the subgroup with $S(1)=1$. 
We discuss this general point later in Section \ref{sec:tte_ps}.

% BB: tried to incorporate the text below in the paragraphs above

% \kr{If we define ``biomarker'' to stand for a clearly defined clinical post-randomization event, then other relevant clinical questions can be fit in this framework as well. As an example, longer survival for responders, as compared with nonresponders, should generally not be used to conclude that response {\it caused} longer survival. , . Here, $S$ would be the response status again at a given landmark time $t^*$ and a scientific question of interest would be asking about the treatment effect in those that would respond under treatment.

% These summary statistics derived from the TGI model are post-randomization data and as such they are potentially again subject to immortal bias, as has been criticized in \cite{mistry_16}. While not explictly mentioned in \cite{han_16}, prediction of a treatment effect based on such TGI metrics, at least implictly, tries to answer a causal question. An estimand of interest could be constructed as follows: let $\beta(Z)$ be a continuous variable quantifying {\it ratio of tumor size at $t^*$ to baseline}. Then, define
% \bea
%   S(Z) &:=& 1\{\beta(Z) < \beta^*\}
% \eea for $\beta^*$ a target threshold. A relevant question would then be the treatment effect in those patients with a low (or high) value of $\beta$  under either treatment, i.e. union of the strata $\{S(1)=1\} \cap \{S(0)=1\}$ and $\{S(1)=1\} \cap \{S(0)=0\}$.

% }

% -----------------------------------------------------
\subsection{Antidrug antibodies for targeted oncology drugs}
\label{ex_ada}
% -----------------------------------------------------

In oncology, an increasing number of targeted anticancer agents and
immunotherapies are of
biological origin \cite{brummelen_16}. These biological drugs may
trigger immune responses that lead to the formation of antidrug
antibodies (ADAs). ADAs may be directed against immunogenic parts of
the drug and may affect its efficacy or safety, or they may bind to
regions of the protein which do not affect safety or efficacy, with
little to no clinical effect \cite{moussa_16}. ADA positivity (ADA+)
is triggered by treatment, appears post-randomization and has the
potential to affect the interpretation of the outcome. It can thus be
considered an intercurrent event in the language of the ICH E9(R1)
guideline. Note that in an RCT it can well be that a biologic drug is
only administered in the test but not the control arm, \textit{i.e.}
by construction ADAs can only form in the intervention arm. To make
things concrete, assume that our outcome of interest $Y$ is again a
time-to-event endpoint, e.g. OS. The intercurrent event $S$ is
occurrence of an ADA at a fixed milestone time point $\tilde{t}$ after
randomization, e.g.  $\tilde{t} = 3$ weeks.

A relevant clinical question for the intercurrent event of ADA-positivity is whether ADA+ patients still benefit from the drug.

%A naive analysis would be to simply looking at subgroups generated by
%observed ADA status, e.g. compare ADA- patients on treatment vs. those on control. This would only lead to valid inference under the
%assumption that ADA status is unrelated to treatment.
%\bb{Should this be: This
%  would only lead to valid inference under the assumption that ADA
%  occurence on treatment is unrelated to the potential event time on
%  control}.
%This assumption is quite implausible here, implying that
%the naive analyses would most likely give biased results.

One way to answer the above clinical questions is to assess the effect
of the randomized treatment in those patients that would be ADA+ under
treatment, \textit{i.e.} we are interested in the treatment effect in the
stratum $\{S(1) = 1\}$. This is the union of the two strata
$\{S(1)=1\} \cap \{S(0)=1\}$ and $\{S(1)=1\} \cap \{S(0)=0\}$ in Table
\ref{tab:principal_stratum}. The effect can then again be quantified
via $U_1$ and $U_0$ introduced in Section~\ref{early_biomarker}. 
% Note
% that by the assumption that ADA+ only develops under treatment the
% stratum $\{S(0)=1\}$ is empty and that data on the endpoint $Y$ {\it
%   does not exist} for patients in the control arm that would be ADA+.\bb{Typo? Data on $Y$ does exist
%   for ADA- patients on the treatment arm...}\kr{Thanks - better
%   now?}\bb{Still not clear to me: I propose to remove the sentence
%   starting with ``Note that by assumption ...''}
% 
% \bb{I would either expand or remove this paragraph. E.g. explain in
%   detail what the estimand of interest is and why this is not targeted
% by the performed analysis.}
% In a simple example of a one-arm trial to assess the impact of ADAs on the efficacy of ipilimumab in metastatic melanoma patients, \cite{kverneland_18} simply compare progression-free and overall survival (OS) in the subgroups of patients that are ADA+ at any time during their time in the trial to those who remained ADA-. This analysis suffers from immortal bias and it is unclear whether the conclusion in the paper (``Development of anti-drug antibodies is associated with shortened survival'') can reliably be drawn.

Enrico et al. \cite{enrico_19} give an overview of the issue of ADAs
in the class of immune checkpoint blockers and re-analyze data of drug
trials, by ADA status. They define ADA-positivity by ``patient has
ever been ADA+ during the observation period''. However, as discussed
in the previous section, naive analyses defining groups through a
post-randomization event will lead to (i) the comparison of
non-comparable population on the treatment groups and (ii) in this
example also to immortal bias (see also e.g. \cite{anderson_83,
  anderson_01, walraven_04, anderson_08} and the discussion later in
Section~\ref{sec:tte_ps}). Here, ADA+ patients were not at risk, and
thus ``immortal'', of experiencing the outcome event between trial
entry and the occurrence of ADA positivity.  How much a causal
conclusion based on the analysis in \cite{enrico_19} is justified is
thus unclear. In the context of TGI metrics discussed in
Section~\ref{early_biomarker} this has been brought up in
\cite{mistry_16} as well.

% Landmark analyses, see e.g. \cite{anderson_08}, would be an option alleviating the immortial bias issue. The landmark method calculates survival estimates from a milestone time point $t^*$ on and statistical inference is performed conditional on patients' landmark ADA status. This eliminates immortal time bias, at the cost of breaking the randomization and having to remove patients with follow-up less than or dying prior to the landmark time. Adjustment for prognostic factors is possible to deal with the first issue, but the implied estimand bears no causal interpretation.

% -----------------------------------------------------
 \subsection{Impact of exposure on OS}
 \label{ex_trough}
 % -----------------------------------------------------

 The {\it Trastuzumab for Gastric Cancer} (ToGA) trial was a 1:1 Phase
 3 RCT comparing chemotherapy vs. chemotherapy + trastuzumab in
 patients with gastric or gastro-oesophageal junction cancer with
 over-expression of the HER2 protein \cite{bang_10}. 584 patients
 entered the primary analysis. A post hoc exploratory analysis of OS
 by trastuzumab exposure in the intervention arm of ToGA was also
 performed \cite{cosson_14}, where exposure was defined as trough
 minimum concentration, $\cmin$, at steady state in Cycle 1. Clearly,
 $\cmin$ is a post-randomization variable. The authors observed that
 patients with $\cmin$ values in the lowest quartile of the $\cmin$
 distribution appeared to have shorter OS duration compared with other
 quartiles. In order to explore whether other (baseline) factors than
 exposure could contribute to the shorter observed OS in the lowest
 quartile group, further analyses of baseline patient characteristics
 by $\cmin$ quartile were performed. The conclusion was that ``...it
 is unclear whether the lower OS is due to low drug concentration or
 to disease burden.''  In a follow-up analysis authors by the FDA in
 \cite{yang_13} evaluated the treatment effect in patients in the
 lowest $\cmin$ quartile (in the chemotherapy + trastuzumab arm) by
 appropriately matching these patients with patients in the
 chemotherapy only arm, to achieve covariate balance for key baseline
 covariates. Although not explicitly described in the potential
 outcomes framework, this approach implicitly targets a principal
 stratum estimand with $S$ being an indicator for $\cmin$ below a
 given threshold after Cycle 1 on the test treatment, so that we are
 again in the situation of Section~\ref{early_biomarker}. This
 analysis together with further exposure-response analyses based on
 the ToGA data then triggered initiation of a fully-powered open-label
 RCT, HELOISE, that evaluated standard vs. high dose trastuzumab
 \cite{shah_17}, for the identified subgroup of patients. This
 case-study illustrates the clinical importance of principal stratum
 estimands might have on a drug development program, even if not
 explicitly mentioned in the \cite{yang_13} paper.

\subsection{Prostate cancer prevention trial}
\label{ex_prostate}
% -----------------------------------------------------

The prostate cancer prevention trial (PCPT, \cite{thompson_03}), a
double-blind RCT, randomized 18882 men aged 55 years or older to
finasteride or placebo. The trial convincingly showed that men
randomized to finasteride had lower rates of prostate cancer. However
in \cite{thompson_03} it was noted that among patients who developed
prostate cancer after randomization, those randomized to finasteride
had a statistically higher risk of high-grade prostate cancer compared
to those randomized to placebo. The question of interest here is
therefore assessing the effect of finasteride on the severity of
prostate cancer among those men who would be diagnosed with prostate
cancer regardless of their treatment assignment, see \cite{lu_13} for
a very nice discussion. Severity was measured using the Gleason score,
an ordered categorical variable that assigned integer values 2-10,
with 10 being the most severe. To make things concrete, $Z$ is the
indicator of being randomized to finasteride, $S$ is the indicator of
getting prostate cancer, and $Y$ is the Gleason score.

Interest thus focuses on the distribution functions of the two
potential outcomes $Y(0)$ and $Y(1)$ in the stratum of those patients
who get prostate cancer irrespective of treatment assignment,
\textit{i.e.} $\{S(0) = 1\} \cap \{S(1) = 1\}$. \cite{lu_13} describe how to
estimate this effect and how to statistically test equality of
distribution functions of the two potential outcomes.

In terms of results, naively looking at the distribution of Gleason
scores in both arms suggests that those who got cancer in the
finasteride arm had higher Gleason scores. This naive analysis however
did not account for potential post-randomization selection bias due to
differences among treatment arms in patient characteristics of cancer
cases or differential biopsy grading associated with
finasteride-induced reductions in prostate volume \cite{lucia_07}. A
subsequent sensitivity analysis based on principal stratification
\cite{shepherd_08} accounting for these two potential sources of
selection bias cast doubt on results from the aforementioned naive
analysis. 
% This analysis based on principal stratification provided
% robust evidence that finasteride did not cause higher Gleason scores among
% those who were doomed to get cancer {\it regardless} of their
% treatment assignment\bb{Shouldn't it be ``provided robust evidence
%   that finasteride does not cause higher Gleason scores''}. 
  Indeed, a
more recent report based on long-term follow-up of PCPT patients
\cite{goodman_19} has concluded that {\it The early concerns
  regarding an association between finasteride and an increased risk
  of high-grade prostate cancer have not been borne out.}

% -----------------------------------------------------
\subsection{Further examples}
\label{further_examples}
% -----------------------------------------------------

Targeting a principal stratum estimand has also been suggested for a
variety of further examples and we sketch and reference some of these
below.

A further application of principal stratum is discussed by
\cite{lou_19b, lou_19a} in the context of clinical equivalence
studies, where intent-to-treat analyses are not considered
``conservative'' and per-protocol analyses are also commonly
performed as primary analysis. Here the authors propose to use protocol adherence for $S$,
and interest is in the stratum of patients that would adhere under
both treatment and control. This corresponds to an estimand which is
in spirit similar to the aims of the per-protocol analysis. In that
direction the ICH E9(R1) guideline in Section A.5.3 also critically
discusses per-protocol analyses, but also argues against handling of
protocol deviations as one intercurrent event (protocol deviations may
not necessarily be intercurrent events, and vice versa), and thus
suggests a more granular handling of the different intercurrent events
that may underlie protocol non-adherence.

Uemura et al. \cite{uemura_19} propose to use principal stratum
estimands for assessing quality of life in face of an intercurrent
event that might happen before the assessment of quality of
life. Typically in this case naive analyses are performed that ignore
the intercurrent event.

In the context of schizophrenia, Larsen and Josiassen \cite{larsen_19}
are interested in the treatment effect on a continuous outcome $Y$ in
patients that would comply if treated with the test treatment,
\textit{i.e.} the effect in the stratum $\{S(1) = 1\}$ in
Table~\ref{tab:principal_stratum}. They propose a new estimator for
this setting.

Akacha et al. \cite{akacha_17} suggest the tripartite estimand
approach for characterizing the treatment effect in the overall
population. They suggest reporting three numbers (i) non-adherence due
to safety (ii) non-adherence due to lack-of-efficacy and (iii) the
effect in adherers. Estimand (iii) corresponds to a principal stratum
strategy, see also \cite{qu_20} for a concrete application of this
approach in a diabetes trial.

In the context of COVID-19 vaccine development a vaccine-induced
dampening of disease severity might be clinically relevant (see
\cite{fda_20_covid}), that is to reduce the {\it severity} of COVID-19
disease among the subset of those who will become infected despite
vaccination.

% Further potential examples to maybe mention in discussion:
% \bi
% \item \cite{jemiai_07}: benefit of vaccine therapy on survival in the stratum who would become infected regardless of treatment with vaccine or placebo (``always infected'' PS).
% \item Embed treatment switching in principal stratum framework. Pre-addendum example from Novartis? \bb{Bjoern to check}.
% \item Duration of response in those who respond on either treatment in onco trials?
% \item Devan's vaccination example.
% \ei

% -----------------------------------------------------
\section{Analysis methods and assumptions}
\label{sec:analysis}
% -----------------------------------------------------

Estimates for principal stratum estimands rely on the validity of
assumptions beyond randomization. In the literature, a variety of
possible assumptions have been suggested and the choice of the most
appropriate particular set of assumptions will depend on the context
of each specific case. The literature on analysis methods for
principal stratum estimands is vast and an extensive review is beyond
the scope of this article. In what follows we provide a selected
overview of commonly utilized assumptions. In addition, we discuss
possible sensitivity analyses in Section~\ref{sec:sens_analysis} and
considerations specific to principal stratum estimands for
time-to-event endpoints in Section~\ref{sec:tte_ps}.

To allow readers to implement some of the analyses presented below, we
have developed a \proglang{R} \cite{r} markdown \cite{xie_18,
  allaire_20}, see Section~\ref{software} for the link. The file
generates an exemplary clinical trial data-set containing potential
outcomes $Y$ and $S$ as well as a categorical covariate $X$ mimicking
the case study in Section~\ref{ex_ada}, and provides explicit code for
some of the analyses described below. Finally, a sensitivity analysis
is sketched.

\subsection{SUTVA and consistency}

Most approaches described rely on the stable unit treatment values
assumption (SUTVA), which entails that (i) the potential outcomes for
any patient do not change with the treatment assigned to other
patients (no interference) and (ii) there are no multiple versions of
treatment. An example where (i) is violated is in the area of
infectious diseases: Depending on the context, whether or not an individual may get infected will
depend on whether other individuals are vaccinated. Part (ii) implies
that treatment needs to be well-defined so that potential outcomes
corresponding to a defined treatment are equal to what is observed in
the trial.  This is sometimes also called ``consistency'' assumption
\cite{vanderweele_13}. In addition, most approaches utilize the fact
that there is an ignorable treatment assignment mechanism, as is common in pharmaceutical RCTs.

\subsection{Identification bounds}
One stream of literature tries to avoid utilizing assumptions. This
means that typically no concrete estimate can be provided but only
identification bounds for the parameters of interest, see for example
\cite{zhang_03, chiba_12}. The estimation problem then focuses on
estimation of these identification boundaries, for which also
confidence intervals can be provided. Often these bounds might be
quite wide and might not provide useful information, but this depends
on the specific data situation. Refinements of boundaries using for
example covariate information were discussed in \cite{grilli_08},
\cite{long_13} and \cite{mealli_13}.

\subsection{Monotonicity and exclusion-restriction}

Two possible ``nonparametric'' assumptions to utilize are the
monotonicity assumption and the exclusion-restriction assumption
\cite{angrist_96}. The monotonicity assumption states that
$S(0)\geq S(1)$ (or alternatively $S(1)\geq S(0)$ depending on the
situation). This means for a patient with $S(0)=0$ observed we would
know that $S(1)=0$, so that the bottom-left stratum in Table
\ref{tab:principal_stratum} would be empty.  This assumption allows to
estimate the principal stratum probabilities. The monotonicity
assumption may in some situations be scientifically very plausible,
but is not verifiable based on observed data.  It however implies that
$P(S(0)=1)\geq P(S(1)=1)$, an assumption that can be assessed in a
RCT.

The exclusion restriction assumption states that for patients in the
strata $\{S(0)=0\}\cap\{S(1)=0\}$ and $\{S(0)=1\}\cap\{S(1)=1\}$ one
assumes $Y(0)=Y(1)$, \textit{i.e.}  there would be no treatment effect
in the strata of those experiencing (or not experiencing) the
post-randomization event under either treatment. Formulated
alternatively, randomization has no impact for those subjects for whom
treatment has no effect on $S$ \cite{joffe_07}.

Note that both assumptions make statements on the relationship of
potential outcomes across treatment and control. As potential outcomes
across treatment and control are never observed jointly, these type of
assumptions are typically not verifiable and can be called
``across-worlds'' assumptions. The naming ``across-worlds'' comes from
the fact that this assumption could only be verified if two
``parallel'' worlds could be observed jointly, in which a patient
would receive control in one world and treatment in the other parallel
world.

In the context of the multiple sclerosis example of
Section~\ref{ms_example} these assumptions together would allow
identification of the estimand of interest. But while a monotonicity
assumption can well be justified based on earlier data, the
exclusion-restriction assumption is not plausible, as the estimand of
interest is the treatment effect in the stratum with no relapses
$\{S(0)=0\}\cap\{S(1)=0\}$.

\subsection{Joint models}

In \cite{frangakis_02} a generic likelihood is described for
estimation of a principal stratum effect. This entails a model for the
outcome given the principal stratum membership: $Y(0),Y(1)|S(1),S(0)$
and additionally the principal stratum membership $S(0),S(1)$ itself
is modeled. Multiplying the likelihoods of both models together,
implies a joint model for $Y$ and $S$. Unobserved potential outcomes
are then treated as missing data in \cite{frangakis_02} and integrated
out to define the likelihood. While covariates are not specifically
mentioned, including them in the model for the principal stratum
membership or the outcome is straightforward. As noted in
\cite{frangakis_02} a unique maximum likelihood estimate generally
does not exist (even asymptotically for ``infinite'' sample size).
Further assumptions are needed, which typically involve statements on
the joint distribution of the potential outcomes across treatment and
control (across-world assumptions). 

In a Bayesian setting, prior assumptions on the model parameters can
be quantified in terms of prior distributions and in this case, as
long as proper priors are used, also posterior inference is
possible. This idea goes back at least to \cite{imbens_97} and was
implemented for estimation in the example in
Section~\ref{ms_example}. There a ``soft'' version of the monotonicity
assumption was used, by specifying an informative prior distribution
for the corresponding principal stratum proportion to be close to
0. This allows for sensitivity analyses through varying the
informativeness of the prior distribution. While the model by
\cite{magnusson_19} in Section~\ref{ms_example} did not include
covariates to model the principal stratum membership, this is possible
(see e.g. \cite{zhang_09}, \cite{hirano_00},
\cite{frumento_12}, \cite{mattei_13}, \cite{mealli_16},
\cite{baccini_17} in a Bayesian setting).

It is often plausible to assume that covariates might influence
principal stratum membership, so that inference will get more
precise. This approach can also be coupled with additional assumptions
like monotonicity and exclusion restriction, which will improve
identification of the underlying inference problem. For example
\cite{jo_09}, and \cite{stuart_15} employ the exclusion restriction
assumption in addition to a parametric assumption in the context of
linear regression, while using the EM-algorithm for ML estimation
(extension to a time-to-event outcome using instrumental variable
approaches is discussed in \cite{mackenzie_16} and
\cite{martinussen_17}). A general challenge, independent of whether
a Bayesian or frequentist inference paradigm is used, is that
inference will depend on the parametric assumptions of the underlying
joint models, as well as the covariates used.

In general, when making a parametric assumption on the distribution of
the data, the parameters are identified by the identification of
parametric mixture models, and an EM-type algorithm can be used for
statistical inference. Note however that the likelihood function may
display pathological behavior, and standard frequentist inference
tools, like bootstrap cannot be used. For further discussion and
references see \cite[Section 2.3.2]{ding_18}.

\subsection{Principal ignorability}
\label{sec:PI}

The last type of assumption we discuss is {\it conditional
  independence}. In the context of principal strata this is often
called {\it principal ignorability} (PI), see \cite{ding_17,
  feller_17} for some recent references. Here, separate models are
specified for $Y$ and $S$, resulting in approaches that are very
similar to propensity score approaches in observational data
analyses. For example, in the early responder example in
Section~\ref{early_biomarker} the estimand of interest was

\bea
P(Y(1) > t|S(1)=1) - P(Y(0) > t|S(1)=1),
\eea

where $Y$ was an event time and $S$ early response. Contrary to
$P(Y(1) > t|S(1)=1)$, estimation of $P(Y(0) > t|S(1)=1)$ is not
straightforward, because $Y(0)$ and $S(1)$ are not jointly observed in
the same patient in a RCT. For patients on treatment that are
biomarker responders, {\it i.e.} $S=1$, the control outcome $Y(0)$ is
unobserved, while for patients on the control arm the biomarker
response status on treatment $S(1)$ is not observed.

%A potential assumption that allows for estimation of principal stratum effects is
%principal ignorability, an assumption similar to the ignorability
%assumption in propensity score analyses of observational
%data.
Now, PI states that conditional on baseline covariates (\textit{i.e.},
confounders) $X$ the $Y(0)$ and $S(1)$ are independent:
$Y(0) \perp S(1)|X$, so the covariates $X$ should include those that
explain both $Y(0)$ and $S(1)$ to the extent that they can be
considered independent. This means once the covariates $X$ are known,
$S(1)$ provides no further information on $Y(0)$ and vice versa,
\textit{i.e.}, the distributional equality \bea p(Y(0) | X, S(1)) &=&
p(Y(0) | X) \eea holds. The benefit from this assumption is that it
allows modeling of $Y(0)$ (or $S(1)$) just based on $X$, the
unobserved outcome does not need to be included in the model. Based on
this, weighting approaches based on propensity scores can e.g. be used
as follows.  First, model the probability that $S(1)=1$ on the
treatment arm depending on $X$, e.g. using logistic regression. Then,
use the predicted probabilities as weights for patients on the control
arm (see \cite{stuart_15, bornkamp_20} among many others). The same
model could also be used in a multiple imputation approach,
\textit{i.e.} imputing $S(1)$ for patients on the control arm. An even
simpler approach may often be standardization
\cite{hernan_20}. Depending on the outcome distribution also plain
regression adjustment for $X$ in the outcome model can be utilized to
estimate a principal stratum effect under the PI assumption. Finally,
matching is also feasible. The propensity score literature extensively
discusses the pros and cons of the different analysis techniques, see
e.g. \cite{austin_11}.

The main assumption of PI is that $X$ contains all variables that
potentially confound $Y(0)$ and $S(1)$ (no unmeasured confounding).
As $Y(0)$ and $S(1)$ cannot be jointly observed, this assumption is
hence across-worlds and not verifiable. Its plausibility needs to be
considered on a case-by-case basis.

One practically relevant and important question is how to decide on
the covariates $X$ to use. Here an important point is to adjust for
all confounders that make the potential outcomes of post-randomization
event occurrence and the final outcome independent. So, formally only
covariates that confound the two outcomes should be adjusted for, that
is, formally one should not include covariates that help predicting
the intercurrent event but have no impact on the outcome. The
discussion on which variables to utilize is very similar to the
discussion in observational data settings, where one tries to find
predictors of both treatment and outcome. A helpful recent overview is
provided by \cite{persson_17}, who study the nonparametric setting.
% \bb{I would propose to remove the Brookhart reference and the sentence
%   below. The paper is also cited in the Persson paper, and the Persson
%   paper is clearer} In the parametric setting for example
% \cite{brookhart_06} provide simulations that show that inclusion of
% covariates that impact outcome, even if unrelated to the intercurrent
% event (treatment for observational data), will lead to an increase in
% efficiency in treatment effect estimates for the outcome model.

While the approach by Larsen and Josiassen \cite{larsen_19} also
utilizes covariates, it utilizes an assumption quite different to
principal ignorability: It appears that utilized covariates are
required to have no effect on the outcome, beyond their effect
mediated via compliance.

\subsection{Sensitivity analyses}
\label{sec:sens_analysis}

Because essentially all analysis strategies targeting principal
stratum estimands require strong assumptions, sensitivity analyses
should be performed, depending on how strong the scientific rationale
for the utilized assumptions is. The proposed sensitivity analyses are
often specific to the specific assumptions utilized so that
consequently a number of different sensitivity analyses approaches
exist, see for example \cite{ding_17, shepherd_11, schwartz_12,
  ding_16} and references cited therein. In a Bayesian approach
sensitivity analyses are often relatively straightforward, as for
example discussed in \cite{magnusson_19}.

In the setting of clinical equivalence trials, Lou et
al. \cite{lou_19a, lou_19b} were interested in the treatment effect in
the patient that adhere under both treatments. They utilize an idea of
\cite{chiba_11} to express the estimand of interest in terms of the
naive per-protocol effect with a bias term.  Estimation of the target
estimand is then done by the naive per protocol effect and varying the
bias term within reasonable bounds in a tipping point analysis.  In
addition they propose to test for equivalence of the proportions of
protocol adherence under both treatments as a co-primary endpoint in
clinical equivalence studies.

Applying methodology initially proposed in \cite{gilbert_03},
\cite{lu_13} in the PCPT example from Section~\ref{ex_prostate} apply
a sensitivity analysis to assess robustness of the naive analysis that
does not account for potential post-randomization selection bias. They
make the distribution function of the potential outcome in the placebo
patients depend on a parameter $\beta$ that can be interpreted as
follows: given someone got prostate cancer in the placebo arm, for a
one-unit increase in Gleason score, the odds that they would have
gotten prostate cancer had they been randomized to finasteride arm
multiplicatively increases by $\exp(\beta)$. Plotting the estimated
relative effect between the two potential outcomes (or the $p$-value
as in \cite{lu_13} if interest focuses on hypothesis testing)
against this parameter $\beta$ then allows an assessment of the
dependency of the conclusion on the amount of selection bias.

For assumptions related to principal ignorability tipping point
analyses can be used to assess the sensitivity of assumptions to the
underlying conclusions. Here $Y(0)$ or $S(1)$ would be used as
potential predictor of $S(1)$ or $Y(0)$ on top of $X$. As the effect
of $Y(0)$ on $S(1)$ or the effect of $S(1)$ on $Y(0)$ cannot be
estimated based on data, these effects would need to be varied in a
sensitivity analysis. Another important sensitivity analysis related
to principal ignorability is to vary the set of confounders $X$
utilized.

\subsection{Special considerations for time-to-event endpoints}
\label{sec:tte_ps}

As discussed earlier, event-driven endpoints require special
considerations, as the primary event in some situations might be a
competing risk to observing the intercurrent event status, potentially
leading to immortal bias when utilizing naive analyses. If the primary
event is death this is obvious, but this situation might also occur
for example when the primary event triggers a stop of the treatment
and the intercurrent event can only happen while on treatment. Then
intercurrent event cannot be observed after stop of treatment.

Assume we are interested in the subpopulation with $S(1)=1$ and the
outcome $Y$ is of time-to-event type, further let $T_S(1)$ be the
potential event time for occurrence of $S(1)$. Then in the situation
described above the event $S(1)=1$ implies $T_S(1) < Y(1)$, so that
implicitly the stratum of interest would be $\{T_S(1) < Y(1)\}$.

When the intercurrent event status $S(1)$ is observed for every unit
at a fixed time $\tilde{t}$, the occurrence of the primary event $Y$
before $\tilde{t}$ would make observation of $S(1)$ impossible so that
observation of $S(1)$ implies $Y(1)>\tilde{t}$. The stratum $S(1)=1$
would then implicitly be defined as
$\{S(1)=1\} \cap \{Y(1)>\tilde{t}\}$.

In both situations the stratum is no longer only described by $S(1)$,
but also by $Y(1)$. As the main comparison of interest is between
$Y(0)$ and $Y(1)$ and the principal stratum itself is now also defined
in terms of the observed outcome $Y(1)$, it becomes more challenging
to find realistic assumptions that would allow estimation of principal
stratum effects.

When interest focuses on $\{S(1)=1\} \cap \{Y(1)>\tilde{t}\}$ it is
easier to find plausible assumptions when $\tilde{t}$ is ``small'',
that is, very few events $Y$ are expected before time
$\tilde{t}$. This could for example be fulfilled in the early
responder, exposure, or ADA example in Sections \ref{early_biomarker},
\ref{ex_ada}, \ref{ex_trough}, when the intercurrent event status can
be identified early. Depending on the specific situation (e.g., if $Y$
measures non-fatal events), in the three examples above it might also
be possible to assess $S(1)$ even after the event $Y(1)$ has already
happened, so that one would not necessarily be in the situation
discussed in this section, where the event and intercurrent events are
competing risks.

This problem is also discussed in some detail in \cite{mattei_20} in
the context of treatment switching, where the event (death in their
considered case) is a competing risk to treatment switch.

% -----------------------------------------------------
\section{Discussion}
\label{discussion}
% -----------------------------------------------------

We believe that there are a number of relevant questions in drug
development that can be formulated as principal stratum
estimands. These are often not related to the primary objective of the
trial, but can still play an important role to characterize how the
drug works in relevant subpopulations defined by different
post-randomization events.

That these type of questions occur in regulatory interactions and are
considered worthwhile to provide guidance upon, can seen from the
anticancer guidance issued by the European Medicines Agency
\cite{ema_17}, which has a dedicated section (7.6.5) on ``Analyses
based on a grouping of patients on an outcome of treatment''. The MS
example discussed in Section~\ref{ms_example} is also available as
Public Assessment Report of the European Medicines Agency
\cite{sipo_epar}. In addition, of course, the ICH E9(R1) estimand
working group considered the principal stratum strategy as important
enough to list it as one of the five intercurrent event strategies.

Questions about the impact of clinical events such as exposure,
response, or safety events like ADA on the outcome of interest have
always been relevant in drug development. However, although often
criticized in the literature, simple analyses such as comparing
subgroups based on a post-randomization event are not uncommon in an
attempt to answer such questions. Causal effects were, at least
implicitly, claimed from such analysis. 
%In an extreme case this may
%lead to initiation of an RCT that is later stopped at a futility
%interim analysis, as discussed in Section~\ref{ex_trough}
Even though the formal idea of principal stratum estimands had been
proposed in the causal inference literature two decades ago, the
explicit uptake of these methods in the drug development community has
so far been low. While there are examples of analyses that
appropriately would target principal stratum estimands (for example
\cite{yang_13}), explicit use of principal stratum estimands has been
limited. Exceptions exist, e.g. to assess efficacy on a post-infection
endpoint in a vaccine trial in \cite{mehrotra_06} (where a principal
stratum estimand has been used for a primary endpoint). We believe and
hope this will change with the principal stratum approach being
prominently mentioned as a strategy in the ICH E9(R1) guideline. The
advantage of adopting the principal stratum strategy for these
questions is that it provides a clear inferential target. Having an
inferential target is crucial to assess the adequacy of assumptions or
specific analyses.

Even more generally, in our experience approaching traditional
analyses with a potential outcome mindset often allows to make
implicit assumptions of traditional analyses more transparent, as in
the example discussed in Section~\ref{sec:pot_out}.

The type of assumptions typically required for identification of
principal stratum estimands are quite strong and usually
unverifiable. While similar type of unverifiable assumptions have long
been used in drug development, for example missing-at-random or
independent censoring assumptions, the impact might be stronger here
as assumptions are not only used to ``impute'' missing responses for a
potentially small subset of the overall trial population, but
depending on the data situation and the type of assumption might drive
inference. However, availability of a clear inferential target though
based on unverifiable assumptions has to be traded off against
``naive'' analyses whose causal interpretation is unclear, if not to
say invalid.

Design considerations can help to make the utilized assumptions more
plausible, for example to explicitly consider at design stage, which
baseline and post-baseline covariates may be associated with stratum
membership, and subsequent collection of these measurements in the
trial. While not feasible in all situations, also cross-over designs
(similar to the run-in period discussed in ICH E9(R1)) may facilitate
estimation of principal stratum estimands under potentially weaker
assumption.

In general we think however that (i) utilized assumptions need to be
motivated by clinical or scientific insights and (ii) that sensitivity
analyses need to be performed for any analysis targeting a principal
stratum estimand.  While sensitivity analyses for certain assumptions,
e.g. monotonicity, have been proposed in the literature and we tried
to review some ideas for further sensitivity analyses in this paper,
we believe there is a need for further developments and practical
guidance in this area.

% -----------------------------------------------------
\section{Data Availability Statement}
\label{software}
% -----------------------------------------------------

The markdown file discussed in Section~\ref{sec:analysis} is available as a github repository: \url{https://github.com/oncoestimand/princ_strat_drug_dev.git}. The direct link to the markdown file is: \url{https://oncoestimand.github.io/princ_strat_drug_dev/princ_strat_example.html}.

% -----------------------------------------------------
\section{Acknowledgments}
\label{ack}
% -----------------------------------------------------

This paper has been written within the industry working group {\it
  estimands in oncology}, which is both, a {\it European special
  interest group ``Estimands in oncology'', sponsored by PSI and
  European Federation of Statisticians in the Pharmaceutical Industry
  (EFSPI)} and a scientific working group of the biopharmaceutical
section of the American Statistical Association. Details are available
on \url{www.oncoestimand.org}. We are grateful for feedback of working
group colleagues, as well as from Kelly Van Lancker and Fabrizia
Mealli on earlier versions of this manuscript. We would also like to
thank two anonymous reviewers for helpful comments.

% \appendix
%
% % -----------------------------------------------------
% \section*{Appendix}
% % -----------------------------------------------------
% \label{sec:appendix}
%
% If, as in Section~\ref{ex_ada}, a treatment effect for a time-to-event
% endpoint is based on comparison of suitable survival functions, then
% the summary measure comparing the two ideally also bears a causal
% interpretation. Importantly, the hazard ratio generally does {\it not} allow for
% such an interpretation, see e.g. \cite{hernan_10} and
% \cite{aalen_15}. An exception to this is when the survival times follow an exponential distribution, in which case the true between-treatment difference in mean log survival times (an unambiguous causal estimand) is equal to the reciprocal of the log hazard ratio.
%
% \citet{bornkamp_20}[Section 2.1] recommend either the
% difference in milestone survival or restricted mean survival as an
% effect measure. \cite{mao_18} refers to the former as ``restricted
% average survival causal effect'' and the latter as ``survival
% probability causal effect'', and both these allow for a causal
% interpretation. For restricted mean survival time see also
% \cite{huang_18, lawrence_19} and references cited therein.

\bibliographystyle{ama}
\bibliography{stat}

\end{document}